\documentclass[prl,twocolumn,amsmath,amssymb]{revtex4}

\usepackage{wasysym}
\usepackage{graphicx}
\usepackage{dcolumn}
\usepackage{bm}

\begin{document}

\renewcommand{\vec}[1]{{\mathbf #1}}
\newcommand{\gvec}[1]{{\boldsymbol #1}}

\title{Kondo effect in carbon nanotube quantum dot in a magnetic field\\}

\author{D. Krychowski}
\author{S. Lipinski}
\affiliation{
Institute of Molecular Physics, Polish Academy of
  Sciences \\  Smoluchowskiego 17, 60-179 Poznan, Poland
} 

\begin{abstract}
The out-of-equilibrium electron transport of carbon nanotube semiconducting quantum
 dot placed in a magnetic field is studied in the Kondo regime by means of the
 non-equilibrium Green functions. The equation of motion method is used. For parallel
 magnetic field the Kondo peak splits in four peaks, following the simultaneous splitting
 of the orbital and spin states. For perpendicular field orientation the triple peak
 structure of density of states is observed with the central peak corresponding to
 orbital Kondo effect and the satellites reflecting the spin and spin-orbital fluctuations.
\end{abstract}

\maketitle

\section{Introduction}

Carbon nanotubes (CNTs) have emerged as a viable electronic
material for molecular electronic devices because they display a
large variety of behavior depending on their intrinsic properties
and on the characteristics of their electrical contacts [1]. These
systems also  form  the powerful tool for the study of fundamental
many-body phenomena.  An example is the observed Kondo effect in
semiconducting carbon nanotube quantum dots (CNTQD) [2,3].  The
long spin lifetimes, the relatively high Kondo temperature  and
the fact that this effect can be seen over a very wide range of
gate voltage encompassing hundreds of Coulomb oscillations [4]
make CNTQDs interesting candidates for spintronic applications.

    The purpose of the present work is to discuss magnetic field
    dependence of the Kondo conductance of CNTQD. Perpendicular
    field couples only to spin and parallel field influences
    both spin and orbital magnetic moments. For vanishing magnetic
    field and orbitally degenerate states the Kondo effect  appears
    simultaneously in spin and orbital sectors resulting in SU(4) Fermi
    liquid ground state with totally entangled spin and orbital degrees of freedom [5].
    Magnetic field breaks the spin-orbital symmetry and  in accordance to the experiment [2,3]
    our calculations show the occurrence of   the multi-peak structure of  the differential
    conductance reflecting the spin, orbital and spin-orbital fluctuations.

\section{Model}
The low energy band structure of semiconducting carbon nanotubes
is orbitally doubly degenerate at zero magnetic field. This
degeneracy has been interpreted in a semiclassical fashion as the
degeneracy between clockwise and counterclockwise propagating
electrons along the nanotube circumference [1]. In the present
considerations we restrict to the single shell and the dot is
modeled by double orbital Anderson Hamiltonian with additional
interorbital interaction:

\begin{eqnarray}
\lefteqn{{\mathcal{H}} = \sum_{k \alpha m \sigma}\epsilon_{k \alpha m \sigma} c^{+}_{k \alpha m \sigma}c_{k \alpha m \sigma}}
\nonumber\\
&&+\sum_{k \alpha m \sigma}t_{\alpha}(c^{+}_{k \alpha m \sigma}d_{m\sigma}+c.c.)\nonumber\\
 && +\sum_{k \alpha m
  \sigma}\epsilon_{m \sigma}d^{+}_{m \sigma}d_{m \sigma}+\sum_{m}U n_{m+}n_{m-}
 \nonumber\\
  &&+\sum_{\sigma \sigma'}U_{12}
  n_{1\sigma}n_{-1\sigma'}
\end{eqnarray}

\noindent where $m = \pm1$ numbers the orbitals, the leads
channels are labeled by $(m,\alpha)$, $\alpha =L,R$.
$\epsilon_{m\sigma}=\epsilon_{0}+\mu_{orb}m h cos(\Theta)+g\sigma
\mu_{B} h$, $\Theta$ specifies the orientation of magnetic field
$h$ relative to the nanotube axis, $\mu_{orb}$  is the orbital
moment. The first term of (1) describes electrons in the
electrodes, the second describes tunneling to the leads, the third
represents the dot and the last two terms account for intra ($U$)
and interorbital ($U_{12}$) Coulomb interactions. Current flowing
through CNTQD can be expressed in terms of the Green functions
[6]:

\begin{eqnarray}
  I_{\alpha} = \frac{\imath e}{2 \hbar}\int_{-\infty}^{+\infty}\frac{d \omega}{2 \pi}
  \sum_{m \sigma}\Gamma_{\alpha m \sigma}(\omega)\cdot G^{<}_{m
  \sigma}(\omega)+\nonumber
  \\+\Gamma_{\alpha m \sigma}(\omega)\cdot f_{\alpha}(\omega)\cdot\ [G^{+}_{m \sigma}(\omega)-G^{-}_{m \sigma}(\omega)]
\end{eqnarray}

\noindent where $G^{<}$,$G^{+}$ and $G^{-}$ are lesser, retarder
and advanced Green functions, respectively, $f_{\alpha}$ is the
Fermi function of $\alpha$ lead and tunneling rate $\Gamma_{\alpha
m \sigma}=2\pi|t_{\alpha}|^{2}\varrho_{\alpha m \sigma}$, where
$\varrho_{\alpha m \sigma}$ is the density of states of the leads.
The total current is given by $I=(I_{L}-I_{R})/2$. The lesser
Green function $G^{<}$ is found using Ng ansatz [7], according to
which the lesser self-energy $\Sigma^{<}$ is proportional to the
self-energy of the corresponding noninteracting system
$\Sigma^{<}(\omega)=A\cdot \Sigma^{<}_{0}(\omega)$, and A can be
found by the Keldysh requirement
$\Sigma^{<}-\Sigma^{>}=\Sigma^{+}-\Sigma^{-}$. The Green functions
are found by the equation of motion method using the
self-consistent decoupling procedure proposed by Lacroix [8].
\\
\section{Results and discussion}

The first point of our numerical analysis is addressed to the
experiment of Jarillo-Herrero et al. [2],  in which  the
conductance of CNTQD for  the  almost parallel field orientation
was examined ($\Theta \simeq 21^{\circ}$). The calculations were
performed with Coulomb interaction parameters $U = U_{12} = 40
meV$, inferred from the size of   Coulomb diamonds.  The addition
energy spectrum indicates that the level spacing of examined
CNTQDs $\Delta\epsilon \simeq 4.3 meV$ [9], what corresponds to
the length of NCT $L \sim 400nm$. The estimated Kondo temperature
is $T_{K} \sim7.7 K$ [2]. Our discussion is based on the single
shell model (1) with the level placed in the centre of Coulomb
valley ($\epsilon_{0} = -20meV$). Such an oversimplified approach,
which gives  only a first crude insight is justified since $\Delta
\epsilon/k_{B}T_{K} \sim6.5$ is large and the higher levels do not
play an important role [10]. To get the experimental value of the
Kondo temperature one has to assume a value of coupling to the
leads $\Gamma = 3.2meV$, which is slightly higher than the
observed broadening of  atomic or Coulomb lines for NCTs examined
by Jarillo-Herrero et al.[2,9].
 The fact that the single level description of the
multilevel systems underestimates Kondo temperature is well known
in literature [10,11]. Orbital moment is estimated from the
average slope between the two Coulomb peaks that correspond to the
addition of the electrons to the same orbital state  and reads
$\mu_{orb} \sim 13 \mu_{B}$ [2]. We focus on the regime, where the
quantum dot is occupied by a single electron. Fig.1a presents the
calculated gray-scale plot of conductance versus magnetic field
and bias voltage for $T = 0.34 K$ compared with the corresponding
experimental plot (inset).  The central bright spot of dimension
determined by $T_{K}$ is the region of spin-orbital Kondo effect.
For vanishing bias and magnetic field the Kondo effect appears
simultaneously in spin and orbital sectors resulting in a SU(4)
Fermi liquid ground state.


\begin{figure}
\includegraphics[width=0.7\columnwidth]{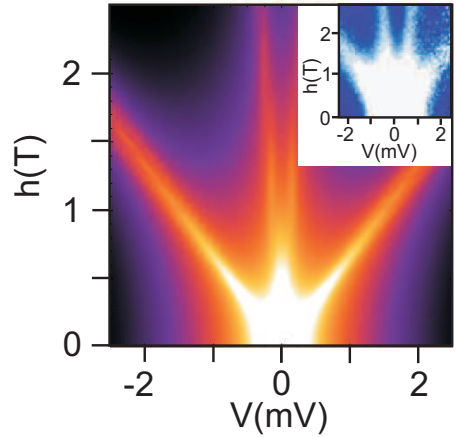}
\caption{Calculated differential conductance $dI/dV$ of CNTQD
versus bias voltage $V$ and magnetic field $h$ in the centre of Coulomb valley  for $T = 0.34 K$. The parameters used are: $U = U_{12} = 40 meV$, $ \Gamma= 3.2 meV$, $\epsilon_{0} = -20 meV$ and $\mu_{orb} = 13 \mu_{B}$. The angle between
the nanotube axis and the field  $\Theta = 21^{\circ}$. Colorscale: $0.1$ to $1.5 e^{2}/h$. Inset shows the corresponding  (V,h)
conductance map obtained from the data of Jarillo- Herrero et
al.[2].}\label{fig1}
\end{figure}

The conductance reaches in the centre  a value  $G = 1.3 \times
e^{2}/h$. Magnetic field breaks the degeneracy and four high
intensity lines appear. A pair of inner lines observed for small
bias corresponds to orbital conserving fluctuations and the outer
lines reflect the orbital fluctuations and simultaneous spin and
orbital  fluctuations. The latter two processes are not resolved
for  the assumed values of $\Gamma$ and temperature. 
 
\begin{figure}
\includegraphics[width=0.7\columnwidth]{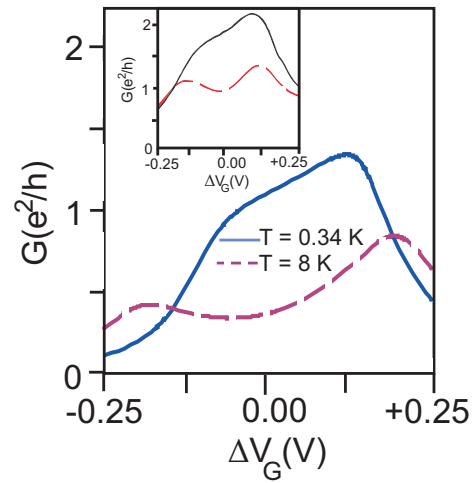}
\caption{Calculated linear conductance $G = dI/dV|_{V\rightarrow0}$ versus gate voltage
            at $T = 0.34 K$ and 8 K for the CNTQD specified by parameters as in Fig. 1a. Inset shows the corresponding curves obtained from the data reported in
[2].}\label{fig2}
\end{figure}

Fig. 1b presents the linear conductance versus gate voltage at $T = 8K$
and $0.34 K$. $\Delta V_{G} =0$ corresponds to the centre of
Coulomb valley. Our calculations reasonably well reproduce the
shape of the dependence but underestimate  all the values of
conductance roughly by a constant value $0.5 \times e^{2}/h$ . The
source of this discrepancy is not clear, but we suggest that a
possible explanation is a neglect  of higher orbital  levels in
our description. Apart from the earlier mentioned  renormalization
of Kondo temperature  they also can cause  a formation of
shoulders in the density of states above the Fermi edge on the
scale much larger than the Kondo temperature [10] and  this might
lead to additional weakly bias dependent contribution to the
conductance. A detailed discussion of the mentioned point will be
given in the following paper [12].

\begin{figure}
\includegraphics[width=0.8\columnwidth]{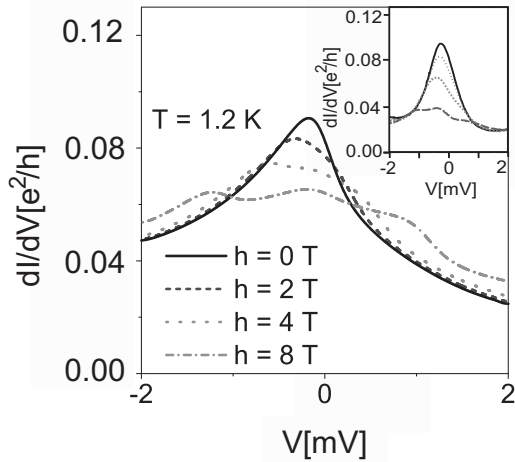}
\caption{Differential conductance of CNTQD in perpendicular
magnetic field for $T = 1.2 K$ calculated for the centre of
Coulomb valley. The parameters used are: $U = U_{12} =15meV$, $\Gamma = 1.25 meV$, $\epsilon_{0} = -8 meV$ with asymmetry of the leads
            $\Gamma_{L}/\Gamma_{R}=6$. Inset shows  the corresponding curves for the same values of magnetic fields obtained from the data of Makarowski et al. [3].
}\label{fig3}
\end{figure}

Now let us turn to the discussion of the  influence of  perpendicular field,
    which breaks only the spin degeneracy. Our numerical analysis describes
    the results of Makarovski et al. [3] for 600nm-long nanotube quantum dot ($\Delta \epsilon \sim  3 meV$).
    The Coulomb parameters  estimated again from the size of Coulomb diamonds are taken as $U = U_{12}
    = 15 meV$ and the orbital level is  placed in the centre of Coulomb valley $\epsilon_{0} = -8 meV$.
$\Gamma$ is taken as the width of orbital or Coulomb peaks.
$\Gamma = 1.25 meV$ , together with the above parameters, well
reproduces  the experimental value of $T_{K}\sim 13 K$.
 The calculated differential
    conductance for several values of magnetic field  is compared with experiment
    on Fig.2. A quasi SU(4) type behavior is still observed in the low field range,
    what reflects in a moderate change of conductance and a single peak structure.
    For higher magnetic fields the spin-orbital Kondo effect SU(4) is
    transformed to SU(2) orbital Kondo effects for each spin orientations seperately.
    This results in the occurrence of the central peak. For bias voltage
    $V =\pm2(\mu_{B}/e)h$ there occur also the satellites induced by tunneling processes which mix different spin channels.

      Summarizing, the present paper provides a simple picture of the
influence of magnetic field on the conductance of carbon nanotube
QDs in the Kondo regime. Although the experiments under
consideration concern multilevel dots our calculations show that
the essence of the transport properties can be inferred from the
effective single shell spin-orbital Kondo physics.

\end{document}